\documentclass[pre,aps,amsmath,amssymb,twocolumn, showpacs,preprintnumbers,floatfix]{revtex4}

\usepackage[T1]{fontenc} 
\usepackage{helvet}      
\usepackage{mathptmx}    

\usepackage{epsfig}
\usepackage{graphicx,color}
\begin{document}

\title{Water permeation through stratum corneum lipid bilayers from
atomistic simulations}

\author{Chinmay Das}
\affiliation{%
School of Physics and Astronomy,%
University of Leeds, LS2 9JT, United Kingdom.}
\author{Peter D. Olmsted}
\affiliation{%
School of Physics and Astronomy, %
University of Leeds, LS2 9JT, United Kingdom.}
\author{Massimo G. Noro}
\affiliation{%
Unilever R\&D, Port Sunlight, Wirral, CH63 3JW, United Kingdom.}



\begin{abstract}
Stratum corneum, the outermost layer of skin, consists of keratin
filled rigid non-viable corneocyte cells surrounded by multilayers of lipids. The
lipid layer is responsible for the barrier properties of the skin.
We calculate the excess chemical potential and 
diffusivity of water as a function of depth in lipid bilayers
with compositions representative of the stratum corneum using
atomistic molecular dynamics simulations. 
The maximum in the excess free energy of water inside the lipid
bilayers is found to be twice that of  water in phospholipid bilayers at the 
same temperature. Permeability, which decreases 
exponentially with the free energy barrier, is reduced by several orders
of magnitude as compared to with phospholipid bilayers. 
The average time it takes for a water molecule to cross the bilayer is calculated 
by solving the Smoluchowski equation in presence of the free energy barrier. 
For a bilayer composed of a  2:2:1 molar ratio of ceramide~NS 24:0, cholesterol
and free fatty acid 24:0 at $300\,\textrm{K}$, we estimate the 
permeability $P=3.7\times 10^{-9}\,
\textrm{cm/s}$ and the average crossing time $\tau_{av}= 0.69\,\textrm{ms}$.
The permeability is about 30 times smaller than existing experimental
results on  mammalian skin sections.

\end{abstract}
\maketitle
\section{Introduction}
The $10-40\,\textrm{$\mu$m}$ thick stratum corneum (SC), 
the outermost layer of the skin \cite{freinkel.skin.01}, 
comprises  rigid keratin filled pancake shaped non-viable cells
(corneocytes) in an extra-cellular lipid matrix. This SC lipid matrix is 
believed to be the main barrier against water loss, with permeability values
three to five orders of magnitude smaller compared to phospholipid
bilayers forming plasma membranes \cite{scheuplein.sc.rev.71}. 
This high permeation barrier
is vital for life to maintain the required ionic concentration inside
the body for proper biochemical reactions. At the same time, controlled 
modulation of the barrier properties has the potential for non-invasive 
drug delivery \cite{prausnitz.04}
and restoration of the barrier function in 
compromised skin \cite{loden.barrier.recovery}.

The SC lipid matrix is conspicuous in having a large fraction of lipids
from the ceramide (CER) family with long and asymmetric acyl tails. 
The other major
components of the lipid matrix are cholesterol (CHOL) and free fatty
acid (FFA) \cite{norlen.sccomp.99, weerheim.sccomp.01}. 
The bricks and mortar model \cite{michaels.sc.brick.75}, one of the widely 
accepted models for the 
SC arrangement, pictures corneocytes as essentially
impermeable bricks joined by the thin lipid matrix forming the mortar phase. 
One of the explanations put forward for the extremely low permeability
of the SC is that the permeating molecules traverse a tortuous route through
the lipid matrix. In this picture, the special chemical structure of 
the lipids plays no important role for the passive permeation. An 
apparent experimental justification for this picture is 
a very large lag time between introducing a radioactive molecule
at one side of the SC and detecting it on the other side. 
With a simple diffusion model, even for water, one needs to invoke a 
path length two orders of magnitude larger
than the physical thickness of the SC layer considered
\cite{potts.sc.perm.91}.
This picture assumes that water does not penetrate the corneocytes. However, 
the corneocytes contain small hygroscopic molecules such as amino acids 
\cite{bouwstra.corneocyte.03}
(collectively  referred to as {\em natural
moisturizing factor} or NMF \cite{jacobson.90, hara.03} ). 
For fully hydrated SC, water diffusivity in the corneocyte is estimated
to be within a factor 2-3 of the diffusivity in 
bulk water \cite{pieper.scdiff.03,kasting.03}. 
So, it is a bit 
contradictory to
consider them as acting as an impenetrable barrier against water transport.

In this work, we limit 
our studies to simulations of hydrated lipid bilayers alone  but include 
the minimal chemical details
appropriate for the SC lipid matrix, and perform a series of molecular
dynamics simulations to probe the permeability of the fully
hydrated bilayers. Ceramide sphingolipids contain a fatty acid
tail attached to a sphingosine motif. The fatty acid tail is highly
polydisperse in length \cite{farwanah.sccomp.cer.05}. Also, there are 
at least 9 different classes of
ceramides in human stratum corneum, with slight variations 
of the head groups, and, in case of ceramide~1, an additional
esterified long fatty acid attached to the longer hydrocarbon tail. 
Free fatty
acids also have a large polydispersity {\em in~vivo} \cite{norlen.sccomp.ffa.98}. 
For simplicity,
we only consider ceramide~NS~24:0 (ceramide 2), with its fatty acid
tail containing 24 carbons. Similarly, the only free fatty acid we consider 
contains 24 carbon atoms. This particular length was chosen to represent the
majority fraction of ceramide and free fatty acid present in the
human stratum corneum \cite{farwanah.sccomp.cer.05, norlen.sccomp.ffa.98}.

CER with their long asymmetric tails control the main distinguishing features
of SC lipid membranes compared to phospholipids
\cite{cdas.bioj.09}. The absence of any large head
group leads to close packing of hydrocarbon tails. We have performed 
simulations of pure CER bilayers, along with more realistic bilayers 
with CER, CHOL and FFA present in either 1:1:1 or 2:2:1 molar ratios. 
Simulation studies of
water permeation for phospholipid bilayers, using very similar force-field
as employed in this study, exist in literature at $350\,\textrm{K}$
\cite{marrink.94}.
To compare and contrast between phospholipid bilayers and SC lipid bilayer,
we also simulated at $350\,\textrm{K}$. For the 2:2:1 composition ratio,
where comparisons are made with experimental results, we have done
additional simulations at $300\,\textrm{K}$.

Our main findings are that the diffusion of water molecules inside
the lipid bilayer is highly anisotropic, and the excess chemical potential
for water is much higher than in typical fluid phospholipid
bilayers at the same temperature. In the presence of the high free energy
barrier, the simple absorption-diffusion picture, with which much
of the experimental results are interpreted in the literature, is no 
longer valid. Instead,
the time taken by a water molecule to cross the bilayer is determined by
the Kramers' first passage time across the bilayer. Our results
suggest that for water, the experimentally determined lag time and thickness of
the stratum corneum can be reconciled without invoking concepts like
tortuosity when the free energy barrier is accounted for correctly.


\section{Computational details}

To calculate the permeability coefficient from simulations, we constrain a 
water molecule at a fixed $z$ distance from the lipid bilayer mid-plane.
Here, the $z$ direction is the direction normal to the bilayer with the
bilayer center of mass being at $z=0$. In this paper, we use the subscript $\perp$ to
denote quantities in the $x-y$ plane of the bilayer.
The average $z$ component of the force on the constrained molecule is 
related to the spatial derivative of free energy \cite{marrink.94, denotter.jcp.98}
\begin{equation}
\frac{d \Delta G(z)}{d z} = - \left< F_z (z) \right>.
\label{eq.fen}
\end{equation}
Here, $\left< \cdots \right>$  refer to averages both over time
and different system replicas.
The value of the local excess chemical potential can be calculated 
through numerical integration from the bulk water phase.
The  auto-correlation of the
force is related to the diffusion coefficient through 
a Kubo relation \cite{marrink.94}
\begin{equation}
D_z(z) = \frac{(R T)^2}{\int_0^{\infty} dt \left<
\Delta F_z (z,t) \Delta F_z (z,0)
\right> },
\label{eq.diff}
\end{equation}
where, $R$ is the universal gas constant, $T$ the absolute temperature and
$\Delta F_z (z,t) = F_z(z,t) - \left< F_z(z) \right>$.
The macroscopic permeability coefficient $P$, the ratio between the flux and concentration
difference ($J = P \Delta c$), is macroscopically 
defined through \cite{marrink.94} 
\begin{equation}
\frac{1}{P} = \int_{-d}^{d} \frac{\exp{\beta \Delta  G(z)}}{D_z (z)} dz,
\label{eq.perm}
\end{equation}
where the integration runs over the bilayer thickness. $\beta \equiv k_B T$,
with $k_B$ being the Boltzmann's constant.
The constraint on the water molecule is only
along the $z$ direction, so the in-plane diffusivity $D_{\perp}$  in the
$x-y$ plane is calculated from
the mean square displacement as a function of time, $\left< \Delta r_{\perp}^2 
\right>(t) = 4 D_{\perp} t$.

\begin{figure}[htbp]
\centerline{\includegraphics[width=3.5in, clip=]{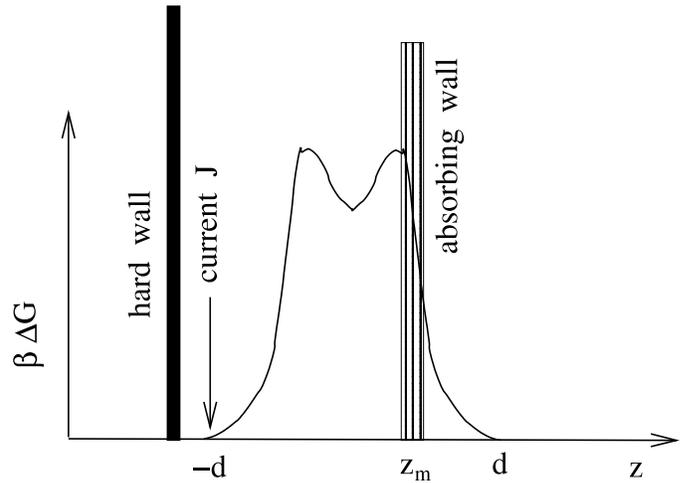}}
\caption{Consideration for calculation of the average
time $\tau_{av}$ for crossing the bilayer. At one side of the bilayer
a steady current is supplied, which is forced to go through the
bilayer because of a hard wall on the other side of
the insertion point. At the maximum of the excess chemical
potential an absorbing wall removes any molecule reaching 
that point.}
\label{fig.tauav}
\end{figure}

To estimate $\tau_{av}$, the average time it takes for a water molecule to
cross a bilayer, we introduce a steady current
$J$ at $z=-d$, at one side of the bilayer (Fig.~\ref{fig.tauav}). 
The number density of water molecules $n(z)$ obeys the Smoluchowski
equation \cite{kramers.review},
\begin{equation}
\frac{\partial n}{\partial t} = \frac{\partial}{\partial z}  
D_z \left(  \frac{\partial n}{\partial z} + \beta n \frac{\partial \Delta G}{\partial z}
\right) + J \delta(z + d) 
.
\label{eq:smol}
\end{equation}
We consider a hard reflecting wall at $z < -d$, where 
$\frac{d \Delta G}{d z} = 0$ (in the bulk $\Delta G \equiv 0$). 
An absorbing
wall is placed at the maximum of the potential $z=z_m$. 
Once steady state is achieved,
the total number of surviving molecules is
\begin{equation}
\int_{-d}^{z_m} n(z) dz = J \tau_{av}.
\label{eq:tkram}
\end{equation}
In the steady state ($\frac{\partial n}{\partial t} = 0$), integrating 
Eq.~\ref{eq:smol}, for $z > d$ leads to
\begin{equation}
n(z) = J e^{-\beta \Delta G(z)} \int_{z}^{z_m} 
\frac{e^{\beta \Delta G(z')}}{D_z(z')} d z'.
\end{equation}
Using this expression in  Eq.~\ref{eq:tkram} gives
\begin{equation}
\tau_{av}= \int_{-d}^{z_m} dz 
 e^{-\beta \Delta G(z)} \int_{z}^{z_m} 
\frac{e^{\beta \Delta G(z')}}{D_z(z')} d z'.
\label{eq:tauav}
\end{equation}
We compute $\tau_{av}$ by numerical integration of functional fits to 
$\Delta G(z)$ and $D(z)$.

All simulations are done with extended ensemble molecular dynamics
at constant temperature and pressure ensemble with GROMACS
molecular dynamics package \cite{gromacs95, gromacs05, gromacs.manual}
with a timestep of 1fs. The interaction parameters are based on
the united atom OPLS force field \cite{jorgensen.ff.88}
with modifications for the nonpolar hydrocarbon groups \cite{chiu.ff.95}
 that accurately reproduce experimental results for lipid molecules
\cite{berger.ff.97}.
Polar hydrogens were included explicitly. The dihedral potentials
in the lipid tails were described by the Ryckaert-Bellemans term 
\cite{ryckaert.ff.75}.
At skin conditions the SC lipids do not fully ionize, so, polar
groups were assigned partial charges chosen from previous simulations
of similar molecules \cite{notman.dmso.07, holtje.fachol.01}. 
The SPC model \cite{spcwater} was used to describe the water molecules.

Nos\'e-Hoover thermostats \cite{nose.thermostat.84,hoover.thermostat.85}
coupled separately to the lipid and water molecules  with a time constant 
of $5\,\textrm{ps}$ were used to control the temperature. 
The Parrinello-Rahman barostat \cite{parrinello.barostat.81, nose.barostat.83}
with time constant $5\,\textrm{ps}$ was used for pressure coupling.
The diagonal components of the compressibility matrix were chosen to be
$4.5\times10^{-5}\,\textrm{bar}$. The off-diagonal components were set
to zero to keep the simulation box orthogonal and standard periodic
boundary conditions were applied in all three directions.  
Electrostatic interactions
were calculated with a group-based cut-off. The cut-offs for both
the Van~der Waals and electrostatic interactions were set to 
$1.2\,\textrm{nm}$. With the small dipole moments involved in these simulations,
the electrostatic interaction becomes negligible at sufficiently small
distance so that the results remain independent of using 
either a group-based cut-off, or 
Ewald summation to take account of interactions with the periodic 
images \cite{cdas.bioj.09}.
All lipid  bonds were constrained with the SHAKE algorithm
\cite{ryckaert.constraint.77}. Rigid SPC water molecules were updated
with the analytic SETTLE algorithm \cite{miyamoto.settle.92}. 

From the final equilibrated configurations from a previous study \cite{cdas.bioj.09}, 
we selected a random water molecule approximately $5\,\textrm{nm}$ 
above the bilayer midplane.
The water molecule was pulled along the
negative $z$ direction at a rate of $0.05\,\textrm{nm/ps}$ by moving
the water molecule $5\times 10^{-5} \, \textrm{nm}$ every (discrete) $\textrm{fs}$
time step. 
Each time the relative $z$ separation between the water molecule 
and the bilayer center of mass  changed by $0.2\,\textrm{nm}$, the 
configuration was equilibrated for
$200\,\textrm{ps}$ with the $z$ separation between the water molecule and
the bilayer midplane kept fixed. The configurations at these steps were 
stored at full precision for further calculations.

We evolved each of these saved configurations, with the selected water 
molecule constrained to be at fixed $z$ separation from the bilayer midplane
for $2\,\textrm{ns}$. 
At each time step, 
the force $F_z$ along the z-direction
on the constrained water molecule
and the 
in-plane displacement of the center of mass of the
constrained water molecule were stored.
Every $100\,\textrm{ps}$ we also store the indices of the atoms
within a distance of $0.4\,\textrm{nm}$ to find out about the local
environment of the constrained water molecule.
The whole procedure was repeated with $15$ different random water
molecules for each composition and temperature investigated.

To calculate $D_z$ from the autocorrelation of $F_z$ (Eq.~\ref{eq.diff}), 
we first estimate the decay time $\tau_c$ of $<F_z(t) F_z(0)>$. 
Typical values of $\tau_c$ are $\sim 0.1\,\textrm{ps}$.
The upper time limit of the integration (eqn.~\ref{eq.diff}) to calculate 
$D_z$  was chosen to be $100 \tau_c$. We calculate the in-plane 
diffusivity $D_{\perp}$ from
long time behavior ($> 0.5\,\textrm{ns}$ ) of the
mean-squared displacements with time origins chosen at intervals
of $1\,\textrm{ps}$ for averaging. 

Data from simulations with different water molecules were 
used together to calculate the average and error estimates
for $F_z (z)$, $D_z(z)$ and $D_{\perp}(z)$.
Because the bilayers studied here are symmetric,
$D_z\, (F_z)$ is an even (odd) function of $z$. 
Using the data from both leaflets we
achieve a  better estimate at a given $z$.
Numerical integration of
$F_z (z)$ from the bulk gives the excess free energy $\Delta G(z)$ 
(Eq.~\ref{eq.fen}).
Finally, permeability is calculated by numerical integration across
the bilayer (Eq.~\ref{eq.perm}) using the calculated $D_z(z)$ and
$\Delta G(z)$.


\section{Permeability of ceramide bilayer}

The asymmetric long chain ceramides are responsible for
much of the distinguishing features of SC lipid structure as
compared to other biologically relevant membranes. In this first part
of this work we concentrate on a pure CER bilayer and contrast our
findings with literature results on phospholipid bilayers.

\begin{figure}[htbp]
\centerline{
\includegraphics[width=3.5in, height=4in, clip=]{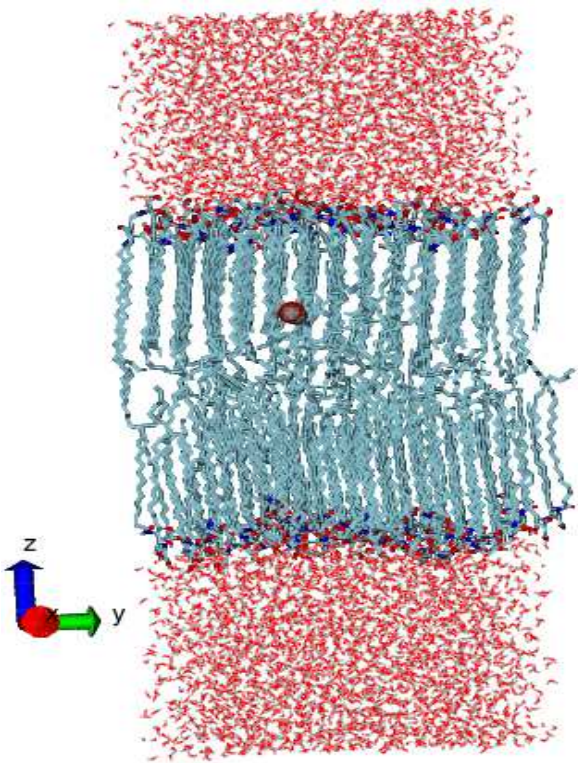}}
\caption{Snapshot of hydrated CER bilayer at $350\,\textrm{K}$
with a constrained water molecule shown as a large sphere.}
\label{fig.confsnapshot.cer}
\end{figure}

Fig.~\ref{fig.confsnapshot.cer} shows a snapshot of CER bilayer 
containing 128 CER molecules and 5250 water molecules with
a water molecules constrained to be at a distance $1.05\,\textrm{nm}$
from the bilayer midplane along the $z$ direction. The long hydrocarbon
tails lead to large nematic order in the bilayer. Once the water molecule
is inside the lipid layer, it faces little resistance in moving along
the $z$ direction, but the motion in the $x-y$ plane is severely restricted.
Because the two tails have large asymmetry, the midplane region is mostly
occupied by atoms from the longer tail of ceramide and there the nematic
order is lower. 

\begin{figure}[htbp]
\centerline{
\includegraphics[width=3.5in, clip=]{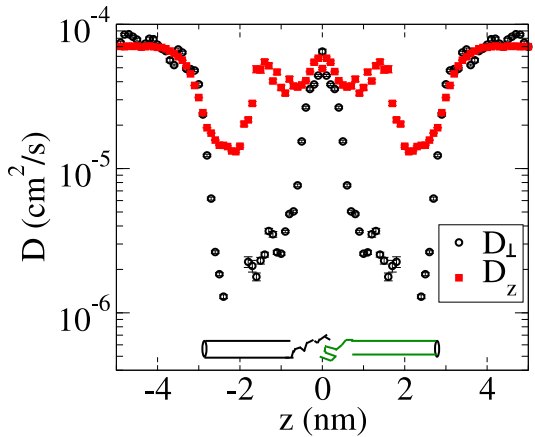}}
\caption{Diffusion coefficients of water as a function of distance
from the bilayer midplane for hydrated CER bilayer at $350\,\textrm{K}$. 
The diffusivity along the normal direction to the bilayer, $D_z$,
is shown as filled square and the in-plane diffusivity in the
$x-y$ plane, $D_{\perp}$ is shown as empty circles. The position of the
head group and the two tails are indicated by the schematic representation
of the ceramide molecule below the plot. We exploit the symmetry about
the bilayer mid-plane to get better averages. Thus data on only one side of $z=0$
is independent.}
\label{fig.diffcer}
\end{figure}

The diffusivity of water as a function of distance from the bilayer
midplane is shown in Fig.~\ref{fig.diffcer}. 
In the bulk water ($|z|  > 4\, \textrm{nm}$), 
both methods lead to very similar value of the diffusivity.
$D_{\perp}$ falls by nearly three orders of magnitude in the ordered
tail region, while $D_z$ is reduced by slightly less than an order. This
agrees with the interpretation of water molecule inside the
ordered region of the bilayer being effectively constrained in 
channels defined by the hydrocarbon tails. Close to
 $|z| \simeq 2 \, \textrm{nm}$, the measured $D_{\perp}$ from
the simulations fall below $10^{-6} \textrm{cm}^2\textrm{/s}$. 
There, the water remains confined in the same lipid
neighborhood for the entire simulation and we can no longer calculate
$D_{\perp}$ with certainty from the mean square displacement. 
Except for this narrow region ($|z| \simeq 2\, \textrm{nm}$), the diffusivity is 
high enough for the water molecule to
explore a large part of the system in the $x-y$ direction within
the run time of a single simulation.
The asymmetry between the two tails lead to a low density molten region at
the bilayer midplane, where both $D_z$ and $D_{\perp}$ approach the bulk
diffusivity and is isotropic. The decrease in the value of $D_z$ inside
the bilayer is comparable to that found in DPPC bilayer simulations
\cite{marrink.94}.
However, in DPPC, $D_{\perp}$ remains comparable to $D_z$ throughout the bilayer.
At the midplane of DPPC bilayer, because of large free volume,
the diffusivity was found to be almost twice that of bulk water.
In ceramide, both $D_z$ and $D_{\perp}$ approach the bulk water value
at the midplane, but remain less than it. 
The asymmetry in the
two tail lengths allow the CER bilayer to have a much larger local density
at the midplane region than the DPPC bilayer, because of  partial 
interdigitation. This limits the diffusivity to a value lower than
that in the bulk.

\begin{figure}[htbp]
\centerline{\includegraphics[width=3.5in, clip=]{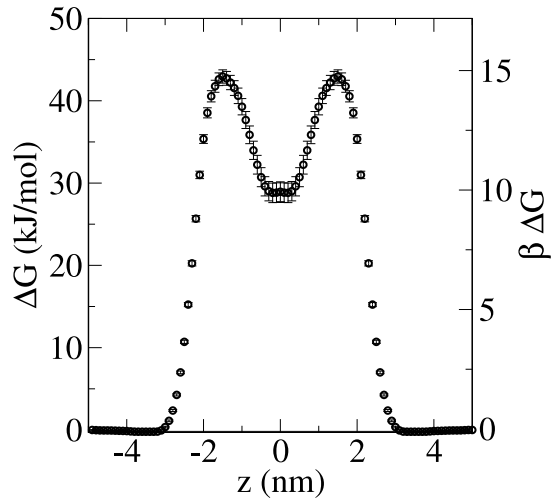}}
\caption{Excess chemical potential for a water molecule 
inside a CER bilayer at a given depth at $350\,\textrm{K}$. 
$z=0$ is the bilayer mid-plane.} 
\label{fig.fencer350k}
\end{figure}

From the average force measured as a function of the depth, we
numerically compute the excess chemical potential $\Delta G$ 
of a water molecule as a function of $z$
(Fig.~\ref{fig.fencer350k}). The chemical potential
rises steeply on entering the lipid bilayer, reaching a maximum
($\sim 43\,\textrm{kJ/mol}\,\simeq 15 k_B T$)
at a distance of $\sim 1.5\,\textrm{nm}$ from the bilayer midplane. 
At the midplane itself, the excess chemical potential drops to
a value of $\sim 29 \, \textrm{kJ/mol} \, (\sim 10 k_B T)$. The maximum
in the excess chemical potential is about twice as high as that measured 
from DPPC bilayer simulation at the same temperature \cite{marrink.94}.

The permeability decreases exponentially with the excess chemical
potential. Our estimate for the permeability for CER bilayer
at 350K, $P = 1.12 (\pm 0.06) \times 10^{-8} \, \textrm{cm/s}$,
is approximately five orders of magnitude smaller than 
the DPPC bilayer permeability of $7 (\pm 3) \times 10^{-2} \, \textrm{cm/s}$
\cite{marrink.94}
at the same temperature. This is mainly due to
the much larger free energy barrier and partially due to 
reduced diffusivity compared to DPPC bilayer.

Using the excess free energy and diffusivity, numerical integration of
Eq.~\ref{eq:tauav} leads to the average crossing time for ceramide bilayer
$\tau_{av} = 3.3\times 10^{-4}\,\textrm{s}$.

\section{Model stratum corneum lipid bilayers}

\begin{figure}[htbp]
\centerline{
\includegraphics[width=3in, height=3.85in, clip=]{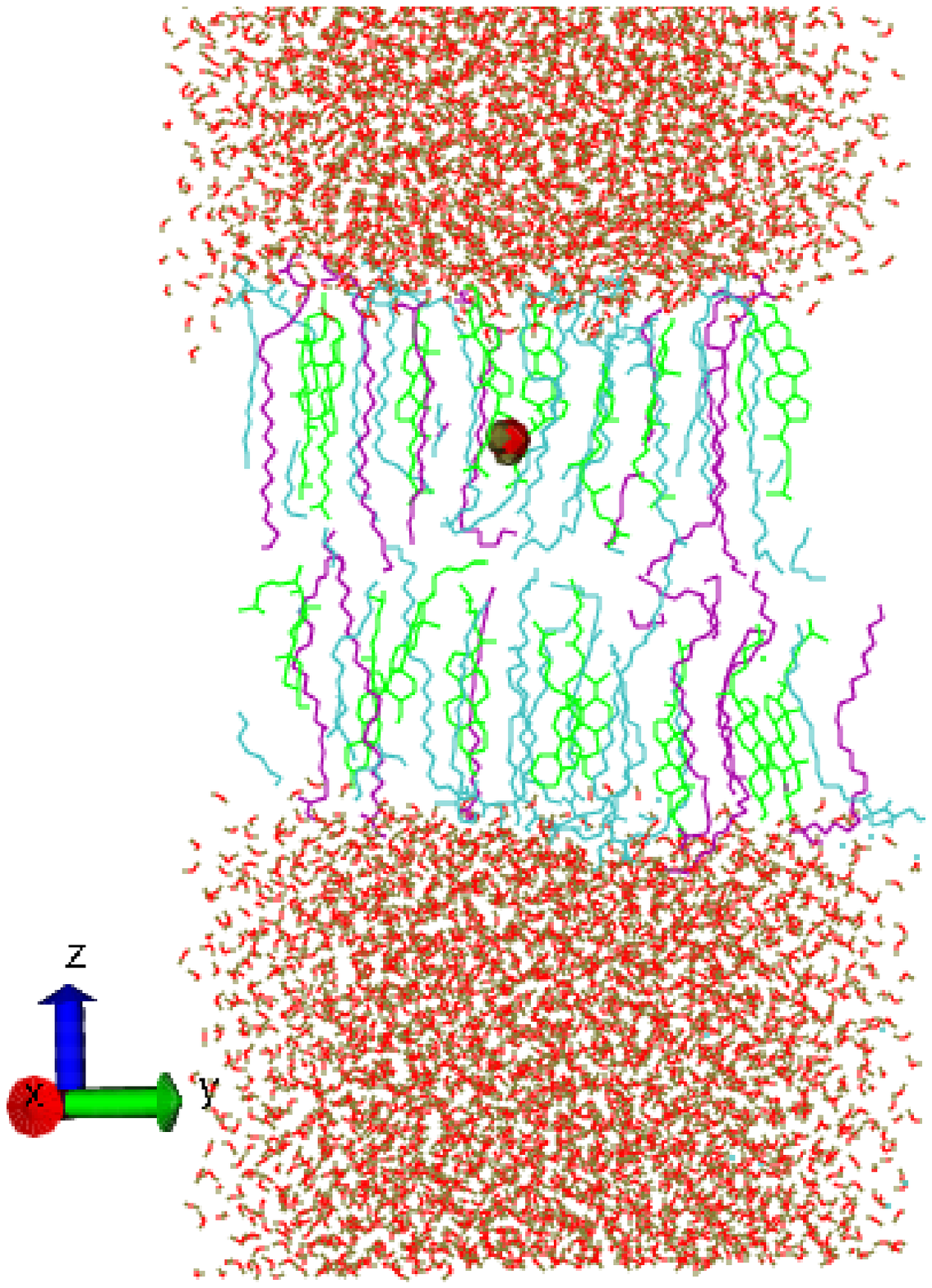}} 
\centerline{
\includegraphics[width=3in, height=3.85in, clip=]{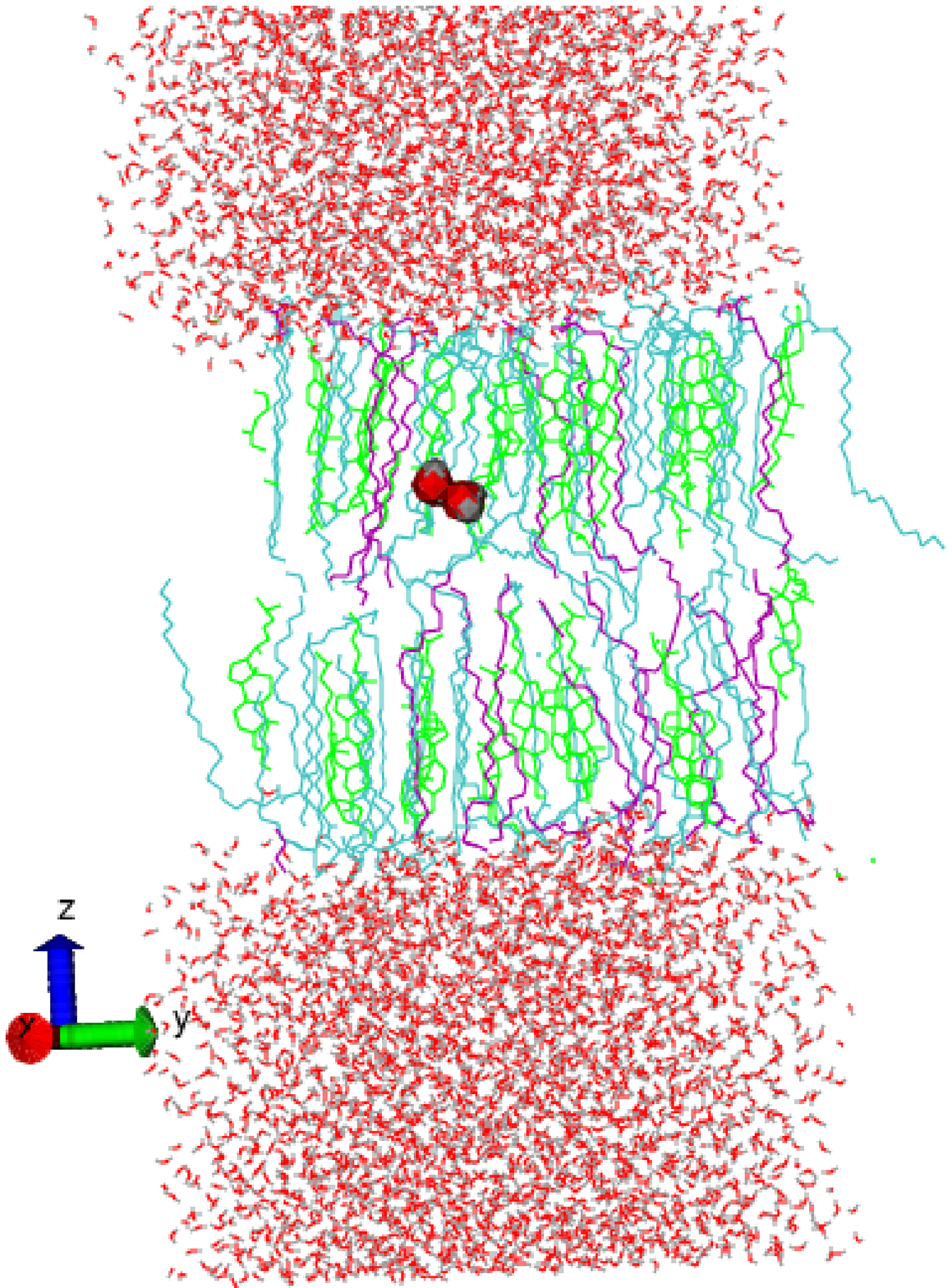}}
\caption{Snapshots of bilayers at $350\,\textrm{K}$ containing a 
2:2:1 molar ratio of 
CER, CHOL and FFA. Part of the lipids were stripped along
the $x$ axis to show the molecular arrangement near the water
molecules (shown as large spheres). The CER, CHOL and FFA molecules are
colored as cyan, green and maroon respectively. In the top panel, only
the constrained water enters the bilayer and in this particular frame is
contained in the free volume created by shorter CHOL molecules.
In the bottom panel the constrained water molecule facilitates entry of another
water molecule in the lipid bilayer. }
\label{fig.confsnapshot.2x2x1_350}
\end{figure}

To provide some understanding for experiments with biological 
skin sections,
we consider two ternary mixtures of CER, CHOL and FFA with 1:1:1 and 2:2:1
molar ratios. The 2:2:1 composition is considered to be representative of 
the {\em in~vivo} composition \cite{norlen.sccomp.99}. 
For this composition, besides $350\,\textrm{K}$, we perform
simulations at $300\,\textrm{K}$, close to the physiological temperature. In this
section we concentrate on the 2:2:1 mixture and consider the results on
the 1:1:1 mixture in Table~\ref{tab:res}. 

CHOL reduces the local 
nematic order in the three component bilayers. Also, to fit the rather bulky 
CHOL molecule, the bilayer needs to create more free volume. 
This effect
on the free volume is opposite to that in phospholipid, where
cholesterol was found to reduce the available free volume in DPPC
bilayer \cite{falck.jcp.04}.
In the simulations with CER bilayers only the constrained water
molecule enters the bilayer, while  in the three component system
another water molecule occasionally joins the constrained water molecule 
(Fig.~\ref{fig.confsnapshot.2x2x1_350}). This
reduces the energy due to favorable hydrogen bonds between the
two water molecules. At the same time, especially close to
the CHOL molecules, large enough free volume is available
to accommodate the two water molecules.

\begin{figure}[htbp]
\centerline{\includegraphics[width=3.5in, clip=]{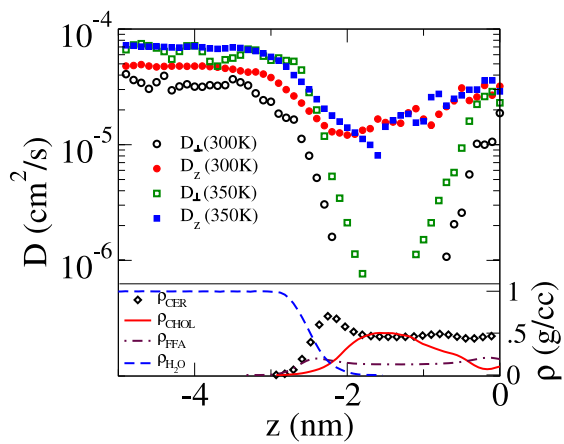}}
\caption{Top panel: Diffusion coefficients $D_z$ (open symbols)
and $D_{\perp}$ (filled symbols) for water molecule as a function of
distance from the bilayer midplane of containing a 2:2:1 molar ratio
of CER, CHOL and FFA respectively at $300\, \textrm{K}$ (circles)
and at $350 \, \textrm{K}$ (squares). 
Bottom panel: Mass densities as a function of $z$ at $300\,\textrm{K}$. }
\label{fig.diffmixed}
\end{figure}

Fig.~\ref{fig.diffmixed} shows the water diffusivity as
a function of $z$ for the 2:2:1 bilayer. 
At the bottom
of the figure, we show the local mass density of the 
three components at $300\,\textrm{K}$. 
FFA follows the CER distribution closely, while
CHOL prefers to stay just below the CER head group, finding favorable
hydrogen bonding with CER. 
Water diffusivity in the bulk scales normally with temperature,
varying from $7.5 \times 10^{-5}\, \textrm{cm$^2$/s}$ at $350\,\textrm{K}$
to $4.5 \times 10^{-5} \,\textrm{cm$^2$/s}$ at 
$300\, \textrm{K}$ \cite{mark.spc.diff}.
At $300\, \textrm{K}$, $D_z$ and $D_{\perp}$ approach each other only
near $|z|\simeq 5 \,\textrm{nm}$, signifying considerable ordering of the
water molecules in the bulk liquid close to the bilayer.
Inside the bilayer the diffusivity is much less affected by the
temperature, because the tail ordering and free volume do not
change by much for this system in this temperature range. The range of $z$,
in which the water molecule is essentially confined in the same
lipid neighborhood during the entire simulation, is larger
than for the pure CER bilayer.

\begin{figure}[htbp]
\centerline{\includegraphics[width=3.5in, clip=]{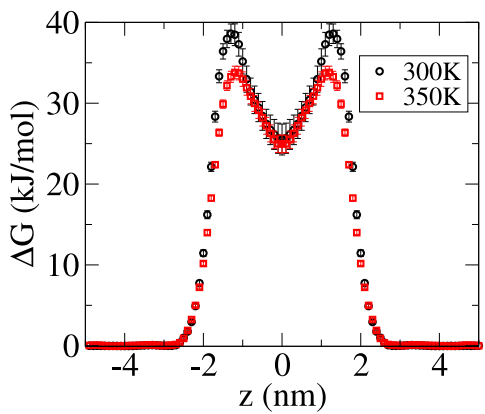}}
\caption{Excess chemical potential for a water molecule 
inside a bilayer containing a 2:2:1 molar ratio of 
CER, CHOL and FFA.}
\label{fig.fenmixed}
\end{figure}

The maximum in the excess chemical potential (Fig.~\ref{fig.fenmixed})
is lower than in the pure CER bilayer ($34 \, \textrm{kJ/mol}$ at $350 \, 
\textrm{K}$ and  $38 \, \textrm{kJ/mol}$ at $300 \, \textrm{K}$). 
At $350\,\textrm{K}$, the permeability of 2:2:1 bilayer 
$P = 1.30 (\pm 0.01) \times 10^{-7} \, \textrm{cm/s}$,
is about one order higher compared to CER bilayer.
Permeability drops by nearly two orders of magnitude on reducing
the temperature to $300\,\textrm{K}$, giving,
$P = 3.67 (\pm 0.02) \times 10^{-9} \, \textrm{cm/s}$.
The average crossing time for the 2:2:1 bilayer increases from
$0.2\times 10^{-4}\,\textrm{s}$ at $350\,\textrm{K}$
to $6.9\times 10^{-4}\,\textrm{s}$ at $300\,\textrm{K}$.





\section{Comparison with experiments}

\begin{table}
\begin{tabular}{|c|c|c|c|c|c|l|}\hline
T & molar ratio &$D_{z, min}$ &  $\Delta G_{max}$ & $\beta \Delta G_{max}$  & P & $\tau_{av}$ \\
(K) &{\tiny CER:CHOL:FFA} &$(10^{-5}\,\textrm{cm$^2$/s})$ & (kJ/mol) &  & (cm/s) & (ms) \\ \hline
350 & 1:0:0 & 1.3 &42.7 & 14.7  &$1.1 \times 10^{-8}$ & 0.33 \\ 
350 & 1:1:1 & 1.1 &34.6 & 11.8  &$8.2 \times 10^{-8}$ & 0.03 \\ 
350 & 2:2:1 & 1.0 &33.8 & 11.6  &$1.3 \times 10^{-7}$ & 0.02 \\ 
300 & 2:2:1 & 1.1 &38.5 & 15.3  &$3.7 \times 10^{-9}$ & 0.69 \\ \hline
\end{tabular}
\caption{Summary of the main  results from the simulations.}
\label{tab:res}
\end{table}

In table~\ref{tab:res}, we summarize the main results for the CER bilayer and
the ternary mixtures. 
The midplane density for the  2:2:1 bilayer at $350\,\textrm{K}$ is 
$\sim 0.7\,\textrm{g/cc}$ \cite{cdas.bioj.09},
comparable to that of liquid hexadecane 
$\sim 0.73\,\textrm{g/cc}$ \cite{bondi.jpc.54}.
The excess free energy of water in hexadecane is $\sim 25\, \textrm{kJ/mol}$ 
\cite{schatzberg.jpc.63}, which agrees with the excess chemical potential at
the bilayer midplane (Fig.~\ref{fig.fenmixed}). 
The strong nematic order induces much larger density close to the
head groups, with average density for the SC bilayers being 
$\sim 0.94\,\textrm{g/cc}$ \cite{cdas.bioj.09}.
This reflects in the maximum in free energy being $\sim 40\,\textrm{kJ/mol}$.
Arrhenius plots of the temperature-dependent permeability from
human\cite{scheuplein.sc.rev.71} and porcine\cite{golden.biochem.87} stratum 
corneum suggest an activation energy $60\,\textrm{kJ/mol}$.

Experiments on human \cite{blank.sc.perm.84}
and porcine \cite{potts.sc.perm.91} SC found a permeability of order
$10^{-7} \, \textrm{cm/s}$.  The permeability value at $300\,\textrm{K}$ for 
the 2:2:1 bilayer from our simulations is $3.7\times 10^{-9}\,
\textrm{cm/s}$, which is about a factor of $30$ lower than the experimental 
values. With exponential dependence of $\Delta G$, permeability depends
strongly on the composition considered. In our simulations, we found 
that the 2:2:1 bilayer shows one order higher permeability 
compared to the CER bilayer at the same
temperature. The lipids in the stratum corneum are highly polydisperse
and contain unsaturated lipid tails. The presence of unsaturated fatty acids
and polydispersity will probably introduce greater disorder, reducing the
permeability as compared to the 2:2:1 bilayer considered in this study.
Also, a patch of stratum corneum is unlikely to have a defect free lipid
structure throughout the sample. Confocal laser scanning microscopy
seems to suggests that the stratum corneum shows large variability in
permeation over $\mu m$ length scales \cite{schatzlein.sc.confocal.98}. 
If there are defect pathways that offer less resistance than the 
defect free lipid layers, the experimental result on the permeability 
is likely to be dominated by these defects. Both the increased disorder 
due to molecular polydispersity and the presence of defects in the macroscopic 
sample will increase the permeability above that of a single perfect 
bilayer (as simulated).

In permeation experiments there is a delay between the introduction
of radioactive water vapour at one side of the sample and its first
detection on the other side. This lag-time $\tau_{lag}$ was experimentally
found to be $\sim 8 \times 10^3\,\textrm{s}$ for porcine skin 
at $303\,\textrm{K}$\cite{potts.sc.perm.91}.
In a Fickian diffusion model, this would correspond to a diffusion path length $\delta\simeq\sqrt{D\tau_{lag}}$.
Reasonable values of the diffusivity $D$ lead to a path
length which is much larger than the physical thickness of the 
sample \cite{potts.sc.perm.91}, and hence the interpretation of a tortuous path that avoids the corneocytes. 

Concentrating on the 2:2:1 bilayer at $300\,\textrm{K}$, and using the
bilayer thickness ($\sim 5 \,\textrm{nm}$) and the minimum in $D_z$
($\sim 10^{-5}\,\textrm{cm$^2$/s}$),
the Fickian diffusion picture yields $\tau_{lag} \sim 10^{-8}\,\textrm{s}$.
This estimate is a factor of  $7\times 10^{4}$ lower than the calculated
mean crossing time ($6.9 \times 10^{-4}\,\textrm{s}$) which takes into account the
free energy barrier (table~\ref{tab:res}). Use of $D_z$ and $\tau_{av}$ in
this Fickian diffusion picture will suggest an apparent pathlength that is
$\sim 250$ times larger than the real bilayer thickness, although the molecule 
actually only traversed the bilayer thickness.

\section{Conclusions}

In conclusion, we have calculated the excess chemical potential profile  and
diffusivity for water molecules
in fully hydrated lipid bilayers composed of lipids corresponding to
the stratum corneum lipid matrix. We find that, compared to 
phospholipids simulated with very similar force fields  and at the same temperature 
as in this study, the SC bilayers show nearly twice as high a chemical 
potential barrier against water permeation. The high degree of correlation of the
free energy profile with the lipid density profile suggests that
the main reason for this large free energy barrier is because
the ceramide lipids lead to dense packing of the acyl tails.
Water diffusivity across the bilayer does not change drastically
as compared to phospholipids. High degree of tail ordering ensures that
once the water molecule is inside the hydrocarbon region, it can
move without large extra free energy cost along the chain-orientation
direction. However, diffusivity in the perpendicular direction
(in-plane diffusivity) is decreased by two orders of magnitude.

The permeability coefficients from our simulations are much smaller than
the experimental results. 
In our analysis, we showed that the high free energy barrier necessitates
the use of a first passage time to estimate the lag time, and neglect of the barrier
leads to an apparent pathlength that is much larger than the physical thickness of
the sample. Our results suggest that, for water permeation, we do not need
to invoke impermeable corneocytes.  

It is possible to discern the importance of the free energy barrier on the lag time 
from experiments at different temperatures. The Kramers' mean first passage formalism
suggests that the temperature dependence of the lag time will be Arrhenius like:
$\ln (\tau_{lag}) \sim 1/T$, while a simple diffusive picture 
suggests $\tau_{lag} \sim 1/T$. We are not aware of any measurements which
looked at this temperature dependence.

\section*{Acknowledgments}
This work was supported by Yorkshire Forward through the grant YFRID Award
B/302. CD acknowledges SoftComp EU Network of Excellence for financial
support and computational resources. The authors thank Jamshed Anwar, 
Simon Connell, Michael Bonner, Andrea Ferrante, Alex Lips, 
Robert Marriott, Khizar Sheikh, and Barry Stidder for 
useful discussions.


\providecommand*{\mcitethebibliography}{\thebibliography}
\csname @ifundefined\endcsname{endmcitethebibliography}
{\let\endmcitethebibliography\endthebibliography}{}

\end{document}